\documentclass{article}
\usepackage{hiph-art}
\usepackage{amssymb,epsfig}
    \newcommand{\pathnow}{}

\newcommand{\nc}{\newcommand}
\nc{\rf}[1]{Fig.\,\ref{#1}}

\nc{\beq}{\begin{equation}}
\nc{\eeq}{\end{equation}}
\nc{\beqa}{\begin{eqnarray}}
\nc{\eeqa}{\end{eqnarray}}
 \nc{\beql}[1]{\begin{equation}\label{#1}}
\nc{\req}[1]{Eq.\,(\ref{#1})} 
\volnumber{XX} \issuenumber{Y} \edyear{2006}                             
\frompage{1} \topage{8}                                              
\recrevdate{4 November 2005}                                             
\def\rightmark{Hadronization of Expanding QGP}
\def\leftmark{J. Rafelski and J. Letessier}
 
\title{Hadronization of Expanding QGP} 
\authors{ 
{Johann Rafelski$^{1,a}$ and Jean Letessier$^{2,b}$ %
\index{Rafelski, J.} 
\index{Letessier, J.} 
}\\[2.812mm]
{\normalsize
\hspace*{-8pt}$^1$ Department of Physics, University of Arizona,\\ Tucson, AZ 85718, USA\\[0.2ex] 
\hspace*{-8pt}$^2$ Laboratoire de Physique Th\'eorique et Hautes Energies\\ 
Universit\'e Paris 7, 2 place Jussieu, 75251 Cedex 05, France
}}
 
\abstract{We discuss how the dynamics of an exploding hot fireball of 
quark--gluon matter impacts the actual phase transition conditions between 
the deconfined and confined  state of matter. We survey the chemical conditions
prevailing at hadronization.   }
\keyword{QGP, strangeness, supercooling, hadronization}

\PACS{12.38. Mh,24.10. Pa,25.75.-q }
 
\makeindex
\begin{document}
 
\maketitle

\section{Introduction}\label{intro}
 It is an open  question  if,  within the short
time  available in a laboratory heavy ion
collision  experiment, $10^{-22}$--$10^{-23}$~s, 
the confined color frozen nuclear phase 
can melt and turn into the deconfined quark--gluon plasma (QGP) state of matter.
Assuming this is the case, we address here the question 
how the dynamical evolution  influences the conditions present at
break-up of the deconfined phase.  
 
\begin{figure}[tb]
\centerline{\hspace*{.20cm}
\psfig{width=6cm,clip=,figure=\pathnow 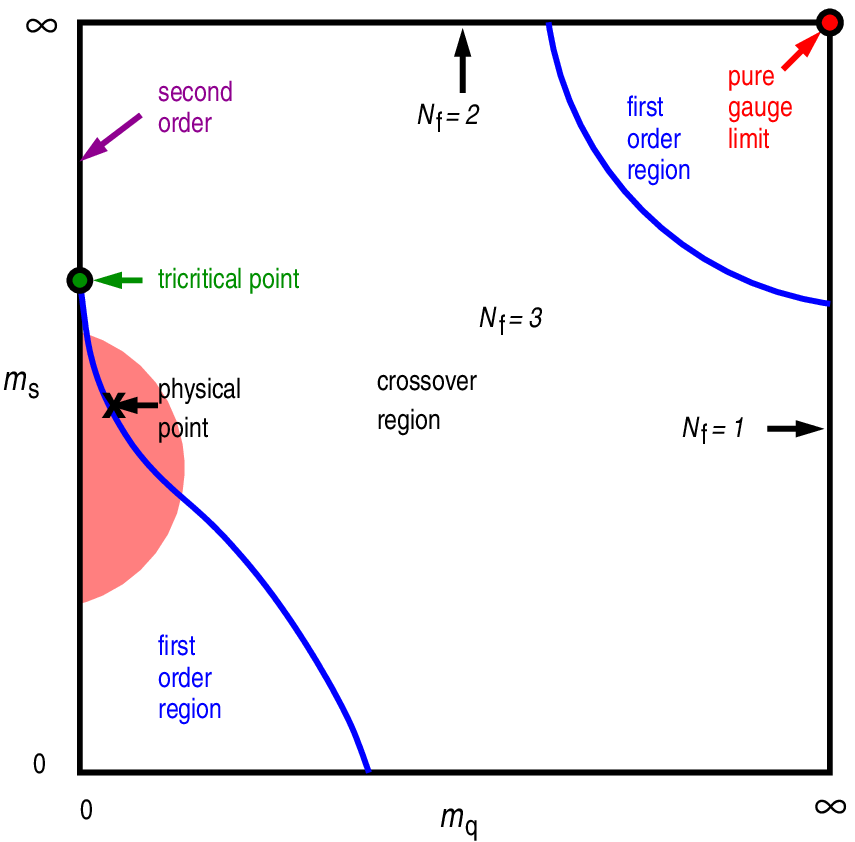}
\hspace*{-.20cm}\psfig{width=6.5cm,height=6cm,figure=\pathnow    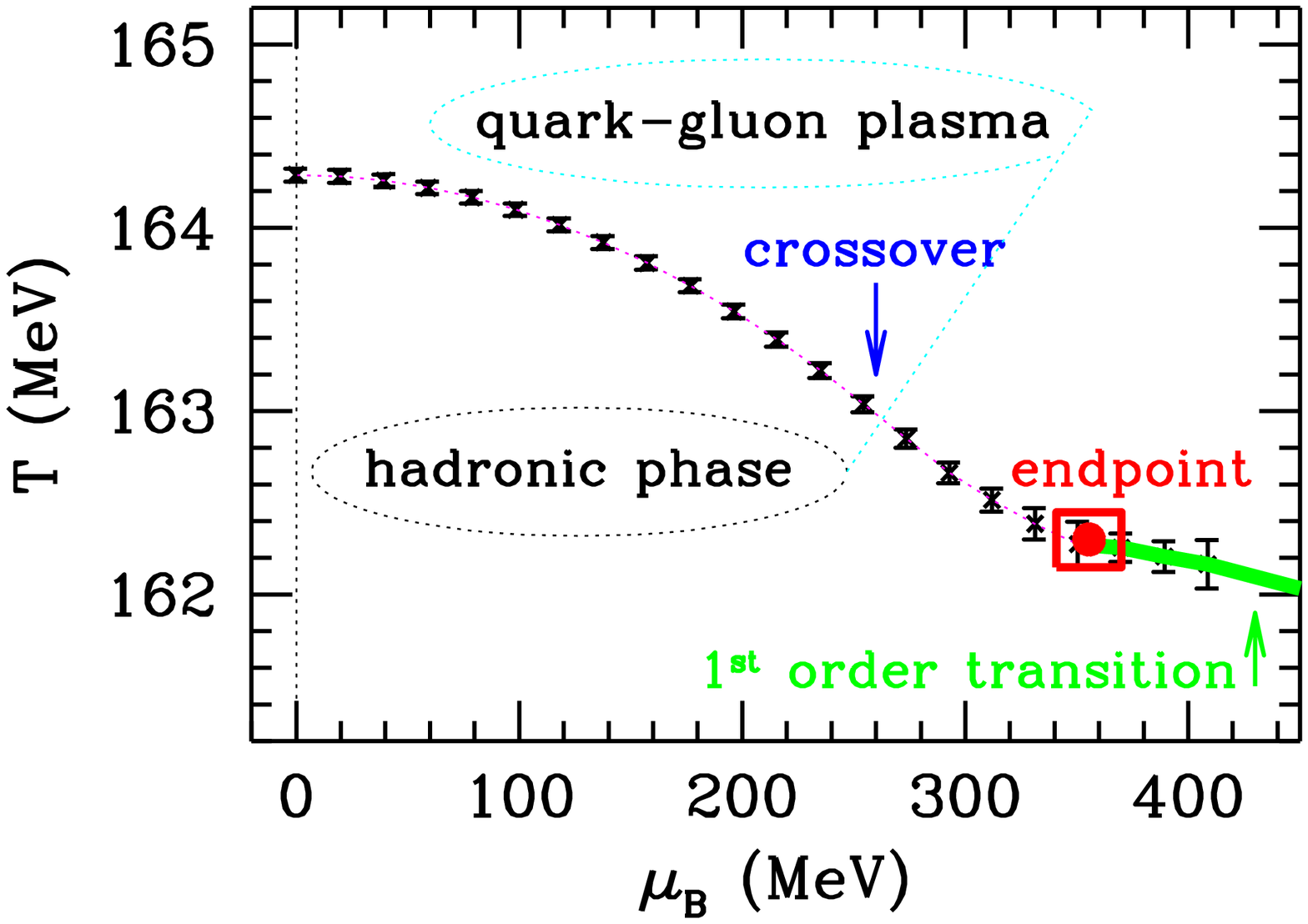}
}
\caption{
\label{2+1Phases}  
QCD thermal phase boundaries, left in the $m_q$--$m_s$ plane~\cite{Peikert:1998jz}, 
and right in the $T$--$\mu_{\rm B}$
plane, for 2+1 flavors~\cite{Fodor:2004nz}.
}
\end{figure}

We will argue that  strangeness production
within  a   fireball of rapidly expanding matter
can   strengthen  the phase transition between QGP and 
hadron gas (HG).
This is of  importance since the QCD thermal state
turns out to remain below a first order transition for a 
physical set of masses of two low mass $m_q/T \to 0$, and one 
semi-heavy quark of mass $m_s/T\simeq 1$.
This is illustrated,
on left, in  Fig.~\ref{2+1Phases}~\cite{Peikert:1998jz}.

 The presence of  a positive quark (baryo-)chemical potential
$\mu_q$ increases the pressure,
and the effect is important when $T\simeq \mu_q$.  
The nearly massless quarks respond more strongly to the finite 
baryo-chemical potential, than do massive hadrons, and that is why
the quark-phase response matters more. For this reason, a
1st-order phase transition is expected to arise at a finite value
of the chemical potential, as is seen on right in 
 Fig.~\ref{2+1Phases}~\cite{Fodor:2004nz,Karsch:2003jg,Bernard:2004je}.
Baryon density, as expressed by the value of the baryo-chemical potential
$\mu_{\rm B}=3\mu_q$, is relatively low at RHIC, where $\mu_{\rm B}\simeq 25$~MeV.
At the much higher energy at LHC,  the expected value of baryo-chemical
potential is of magnitude $\mu_{\rm B}\simeq 1$--$3$ MeV~\cite{LHCPred}.

We recognize, in Fig.~\ref{2+1Phases}, that if 
we could compensate even partially
the  effect of the finite strange quark 
mass, this would suffice to  move the 
physical point  to small and even vanishing
baryo-chemical potential. In chemical non-equilibrium,
the Fermi distribution function takes the form~\cite{cool}: 
\begin{equation} \label{dists}
{d^6N_s\over d^3p\,d^3x} = {\gamma_se^{-\sqrt{m_s^2+p^2}/T} 
                 \over 1+\gamma_se^{-\sqrt{m_s^2+p^2}/T} } .
\end{equation}
For $\gamma_s >1$, there is compensation of the finite 
mass effect~\cite{Letessier:2005qe}. In the limit  $\gamma_s \to e^{m_s/T}$,
the strange quark would play a similar role as a light quark
and we  expect a rather strong first order phase transition 
even at vanishing baryo-chemical potential.

The question we address first is, if such an over-saturation
of the strangeness phase space near to the phase transition
is possible in the dynamical environment we study. We also 
look at any  further dynamical effects associated
with the explosive flow of deconfined matter, and search 
to understand if this   assists   development of the singular
phase behavior.  We close with a discussion of constraints
imposed by approximate conservation of strangeness and entropy in 
hadronization. 

\section{QGP Phase Strangeness Over-saturation}
\label{Tphase}
The   fireball of QGP, created in heavy ion collisions, is initially 
significantly more dense and hot than after its expansion towards 
final breakup condition. This   expansion dilutes the high strangeness 
yield attained in the  initially very dense and hot phase. Contrary 
to intuition, this can result in an over-saturation of the chemical
abundance, even if the initial state is practically strangeness free. 

Considering the yield of strange quark pairs 
 in Boltzmann approximation, at a temperature $T=T_e$, time $t=t_e$,
within the volume $V=V_e$, we have:
\begin{equation}
N_s(t_e)= \gamma_e {2 V_eT_e^3\over \pi^2}
            x_e^2 K_2(x_e), \   x_e={m_s(T_e)\over T_e}.
\end{equation}
We have implied above that the mass of strange quark is     $T$-dependent
as the scale of energy at which its value is determined is in the domain
where a rapid mass change occurs~\cite{CUP}.  
We choose  the  value of  $T_e$ to be the point 
where the system has nearly reached chemical equilibrium abundance
in QGP, with $\gamma_e^{\rm QGP}=1$ (superscript QGP reminds us that
we are considering the deconfined phase).  
 We  assume that the continued expansion  
 preserves entropy as is appropriate for an ideal 
liquid. Since the entropy is governed by essentially massless 
quark--gluon quanta, this implies that $VT^3=$ Const.\,.  
Model calculations show that, for typical values of $T_e$,  the change 
in the absolute number of strange quark pairs  has  
essentially stopped for $T<T_e$. Then, for $t_1>t_e$:
\begin{equation}
{N_s(t=t_1)\over N_s(t=t_e)}\simeq 1 =
   \gamma_s(T_1)  {x_1^2K_2(x_1)\over x_e^2K_2(x_e)}.
\end{equation}
Since the function $x^2K_2(x)$ is a monotonically falling function
(see Fig.\,10.1, p197, in Ref.~\cite{CUP}),  
in general  $\gamma_s^{\rm QGP}(t_1)>1$. In the non-relativistic limit 
$m>T$, that is $x>1$:
\begin{equation}
\gamma_s^{\rm QGP}(T_1) = 
{  x_0^2K_2(x_0)  \over x_1^2K_2(x_1)}  > 1 .
\end{equation}

How large can $\gamma_s^{\rm QGP}(T_1)$ be? Should the QGP breakup occur
from a supercooled state with $T_1=140$ MeV, then it is appropriate
to consider $m_s(T_1)/m_s(T_0)\simeq 1.5$--2, with  $T_0\simeq 210$ MeV. 
To convert from temperature dependence to the energy scale dependence, 
we set $\mu\simeq 2\pi T$
and thus, for $m(T=T_0)$, we need an estimate of $m_s(\mu=1.3\,{\rm GeV})$. 
The PDG value is $m_s(\mu=2\,{\rm GeV})= 80$--130 MeV. Thus, 
 assuming $m_s(T_0)\simeq 140$, we find
$\gamma_s^{\rm QGP}(T_1)> 1.5$ . As this example shows, it is possible to  nearly 
 compensate the effect of the  strange quark mass, see Eq.\,(\ref{dists}).

 We  studied quantitatively the possibility of strangeness (over)population 
at RHIC,  prior to the first experimental results 
becoming available~\cite{Rafelski:1999gq}.
We found    $\gamma_s^{\rm QGP}(T_1)<1.2$ .
We believe that even greater values of $\gamma_s^{\rm QGP}$  
can be  arrived at: we considered the 
instantaneous establishment of transverse expansion 
speed, which cut into the lifespan of the QGP phase and thus 
reduced the production yield of strangeness.
 In addition, the
quark mass chosen was significantly above the currently 
preferred value ($m_s=105\pm 25 $ MeV) which has 
further cut into the rate of production of strangeness.
However, the lower value of $m_s$ also couples
strangeness more strongly to the expansion dynamics  which
could help keep it more in chemical equilibrium. 

There are   studies which
found considerably smaller values of  $\gamma_s^{\rm QGP}$ 
at RHIC~\cite{Biro:1993qt,Pal:2001fz,He:2004df}. In part, this is due 
to employment of dynamical QGP equilibration models that do not allow
gluon chemical equilibration, a prerequisite for abundant 
strangeness formation. Rapid gluon equilibration  processes  are presently not fully 
understood.  We assumed rather rapid glue chemical equilibration
with relaxation time shorter than 1.5 fm. 
We further note somewhat unrealistically low values used for the 
coupling constant $\alpha_s$  ---  we use QCD measured value:
$\alpha_s(M_Z)=0.118$ and use two and higher loop evolution to obtain 
the low energy scale values~\cite{impact}.  We believe that 
if not at RHIC, then at LHC, we should
expect at QGP hadronization a substantial phase space 
excess of strange quarks. 

Though $\gamma_s$ is a useful tool to understand how strangeness
can help to facilitate phase transition, it is important to
remember that it is merely a  quantity which relates actual abundance 
of strangeness to the equilibrium abundance,  
\begin{equation}
n_s\simeq \gamma_s n_s^\infty, \quad  n_s^\infty\equiv n_s(\gamma_s=1).
\end{equation}
A more direct    way to see how 2+1 flavors turns into a  3-flavor QCD
is to note that we hope and expect that the dynamics of fireball expansion
could help us to reach the condition, 
\begin{equation}
n_s(T_1)\simeq n_u(T_1)\simeq n_d(T_1),
\end{equation}
even if at point of chemical equilibrium, 
$n_s(T_e)\simeq 0.5 n_i(T_e),\ i=u,d$. 

In physical terms, relative importance of strangeness increases
since strangeness yield is not reduced along with light quark and 
gluon yields during the dense matter expansion.  
Consider the buildup of collective expansion of the 
fireball of matter which requires conversion of thermal energy into  kinetic energy of 
collective  motion.   The   entropy density decreases, but the expansion
assures that the total entropy remains constant, or slightly  increases. 
As the energy is transfered into the transverse expansion, strange quark pair yield, 
being  weaker coupled, remain least influenced by this loss which mainly consumes
light quarks and gluons.  

We thus conclude that  in the event the initial
 conditions present in QGP are 
sufficiently extreme to generate  strangeness rapidly
and abundantly, one can expect 
over-population of strangeness in the 
final QGP breakup. This condition can   facilitate
the  occurrence of a  phase transition at small and vanishing
baryon density.

\section{Explosive Matter Flow}
\label{exp}
The covariant characterization of the Gibbs condition for  the 
force balance between phases $A$ and $B$, $P\equiv P_A-P_B=0$
requires in the relativistic dynamics 
the introduction of the energy--momentum tensor
$T^{\mu \nu}$. In the laboratory rest frame its components are:
\begin{equation}\label{Tmunurest}
 \widehat T^{ij}=P\delta_{ij},\  
  \widehat T^{i0}= \widehat T^{0i}=0,\ 
  \widehat T^{00}=\varepsilon\,.
\end{equation}
where $\varepsilon$ is the energy density. The
latin indices as usual refer to space component $i=1,2,3$
and the wide hat indicates the laboratory frame. 

Gibbs considered
a space-like  surface along which pressure difference had to vanish. This 
surface is invariantly characterized by a  normal four-vector 
$n^\mu=(0,\vec n)$.
We take as the covariant, frame of reference independent
 statement of the Gibbs condition:
\begin{equation}\label{gibbscov}
T^{\mu \nu}n_\mu n_\nu\equiv 
 T_A^{\mu \nu}n_\mu n_\nu - 
 T_B^{\mu \nu}n_\mu n_\nu =0 . 
 \end{equation}
 
We now consider matter subject to expansion flow.  
$\vec v$ is the velocity of the local
matter element, and its 4-velocity is
$u^\mu=(\gamma,\vec v \gamma), \quad  \gamma={1/\sqrt{1-v^2}}$.
 The natural presence of two Lorentz vectors  
$n^\mu, u^\mu $, assures that 
we cannot transform away the effect of motion, the colored state 
pushes against Gibbs surface where the phase boundary is located. 
We recognize this as the hadronization hyper-surface, where the 
final state hadrons are born. 
 
The components of interest in
the energy momentum tensors are the pressure components:
\begin{equation}\label{Tijv}
 T^{ij}=P\delta_{ij}+(P+\varepsilon)\frac{v_iv_j}{1-\vec v^{\,2}}\,.
 \end{equation}
The balance of forces
between the deconfined  QGP and confined hadron phase 
 comprises the effect of the vacuum which confines color.
 This pressure, introduced  in 
bag models of hadrons for the first time, is
traditionally referred to as the bag constant, 
${\cal B}\simeq (0.2 {\rm GeV})^4$~\cite{Letessier:2003uj}.
This  vacuum structure can be represented within $ T^{ij}$ by:
$P_V=-{\cal B},\quad P_V+\varepsilon_V=0 $.
The   vacuum structure component is thus not entering  the 
dynamical flow term, the last term in Eq.\,(\ref{Tijv}).

We obtain from Eq.\,(\ref{gibbscov}):
\begin{equation}\label{gibbsexpl}
 {\cal B}=P_{\mbox{\scriptsize p}}+
      (P_{\mbox{\scriptsize p}}+\varepsilon_{\mbox{\scriptsize p}})
\frac{\kappa  v^2}
       {1-v^{2}}\,,
\quad
\kappa=\frac{(\vec v\cdot \vec n)^2}
              {v^2}\,,
 \end{equation}
where,
$
P_{\mbox{\scriptsize p}}\equiv 
     P_{\mbox{\scriptsize p}}^{\mbox{\scriptsize QGP}}
   - P_{\mbox{\scriptsize p}}^{\mbox{\scriptsize HG}},\ 
\varepsilon_{\mbox{\scriptsize p}}^{\mbox{\scriptsize p}}\equiv 
     \varepsilon_{\mbox{\scriptsize p}}^{\mbox{\scriptsize QGP}}
   - \varepsilon_{\mbox{\scriptsize p}}^{\mbox{\scriptsize HG}}, 
$
are the {\it particle} pressure and energy density components in the 
pressure and energy density, from which  any  vacuum term has been separated. 
For $\vec v\to 0$, the conventional Gibbs condition reemerges,
$P_{\mbox{\scriptsize p}}^{\mbox{\scriptsize HG}}
      =P_{\mbox{\scriptsize p}}^{\mbox{\scriptsize QGP}}- {\cal B}$.

Eq.\,(\ref{gibbsexpl}) describes the pressure of motion of color charged matter against the 
vacuum structure which is pushed out as color cannot exist there \cite{Rafelski:2000by}; 
we can speak of color wind~\cite{Csorgo:2002kt}.
The magnitude of the   flow effect  on the value of temperature of the 
phase boundary  is relatively large. This is due to
a rather large, in comparison to the pressure, energy density.
In presence of a phase transition,  we can expect
that the discontinuity in particle energy is about 6--7 times greater than
the discontinuity in the particle pressure in  Eq.\,(\ref{gibbsexpl}) ---
we recall that only
the total pressure is continuous as expressed by the covariant Gibbs 
condition, Eq.\,(\ref{gibbscov}).

As a consequence,  we expect a significant down-shift in the critical temperature:
the moving colored fields cause  the quark matter to  super-cool by as much as 
$\Delta T=25$ MeV \cite{Rafelski:2000by}. This means that if the 
equilibrium phase crossover  were to occur at $T=165$ MeV, the dynamical 
cross over would be postponed to the value $T=140$ MeV. We do not know
at this time if the collective and rapid outflow of QCD can turn a  cross-over  into 
a phase-transition.
 
\section{Strangeness and Entropy}
\label{StatHad}
Most of entropy of the deconfined state arises in the initial
parton reactions and rapidly following thermalization. Entropy is
enhanced by color bond breaking, and presence of gluons~\cite{Letessier:1992xd,cool}, 
it is a probe of the early
evolution of the QGP phase. Subsequently, gluon fusion reactions form strangeness~\cite{RM82}.
Both strangeness and entropy are nearly conserved in
the evolution towards hadronization
and thus the final state hadronic yield analysis value for $s/S$ is closely
related to the thermal processes  in the fireball at $\tau\simeq 1$--2 fm/c. 
We believe that, for reactions in which the system approaches strangeness
equilibrium in the QGP phase, one can expect a prescribed ratio of 
strangeness per entropy, the value is basically the ratio 
of the QGP degrees of freedom.

We estimate the magnitude of $s/S$ deep in the QGP phase, considering  
the hot   stage of the reaction~\cite{Rafelski:2004dp}. For  an   equilibrated 
non-interacting QGP phase with perturbative properties: 
\begin{equation}\label{sdivS}
{s \over S}   = 
\frac{ (3/\pi^2) T^3 (m_{  s}/T)^2K_2(m_{  s}/T)}
  {(32\pi^2/ 45)  T^3 
    +n_{\rm f}[(7\pi^2/ 15) T^3 + \mu_q^2T]} 
 \to
  0.027   {  f(\alpha_s) \gamma_s  \over 
  {0.38 \gamma_{\rm G} + 
         0.12 \gamma_s +
         0.5\gamma_q   }},
\end{equation} 
where to obtain the last expression we used for the number of flavors $n_{\rm f}=2.5$,
and  for insignificant baryon density, we neglected the $\mu_q$ term in denominator. 
The numerical value of the coefficient follows  for  $m_{  s}/T=1$.  
All $\gamma_i$ refer, here, to QGP phase. $f(\alpha_s)$ is an unknown factor that 
accounts for the  interaction effects ---  
these maybe canceling, with  $f(\alpha_s)\to 1$ .  
Seen the dependence of \req{sdivS} on $\gamma_i$, we expect to see a gradual 
increase in $s/S$ as the QGP source of 
particles approaches chemical equilibrium with increasing 
collision energy   and/or  increasing number of participating nucleons. 

Since the  ratio $s/S$ is established early on in the 
reaction,    the above relations, and the associated 
chemical conditions  we considered, probe
 the early hot  phase of the fireball. 
How does this simple prediction compare to experiment?
An analysis of experimental data ~\cite{Letessier:2005qe,Rafelski:2004dp},
both as function of energy and 
participant number shows a gradual increase towards the value 
predicted by \req{sdivS}, and thus, one is tempted to conclude
that strangeness nearly saturates at RHIC top energy for most central 
collisions. 
 
We can consider a more direct observable which traces out qualitatively  $s/S$:
  the yield of K$^+$ closely follows 
that of strangeness $s=\bar s$, and the yield of $\pi^+$ is related to the 
total multiplicity $h$, and thus entropy $S$, the experimental
observable of interest is the ratio
K$^+/\pi^+ \propto \bar s/\bar d$ yield ratio~\cite{Glendenning:1984ta}. This 
ratio has been studied experimentally as function of $\sqrt{s_{\rm NN}}$~\cite{Gaz}  
and a pronounced `horn' structure  arises at relatively low reaction energies,
see  \rf{Kpisqrt}. Moreover there seems to be a raise in this ratio after a
dip at intermediate energies.

\begin{figure}[htp]
\centerline{
 \psfig{width=6.7cm,  figure=\pathnow 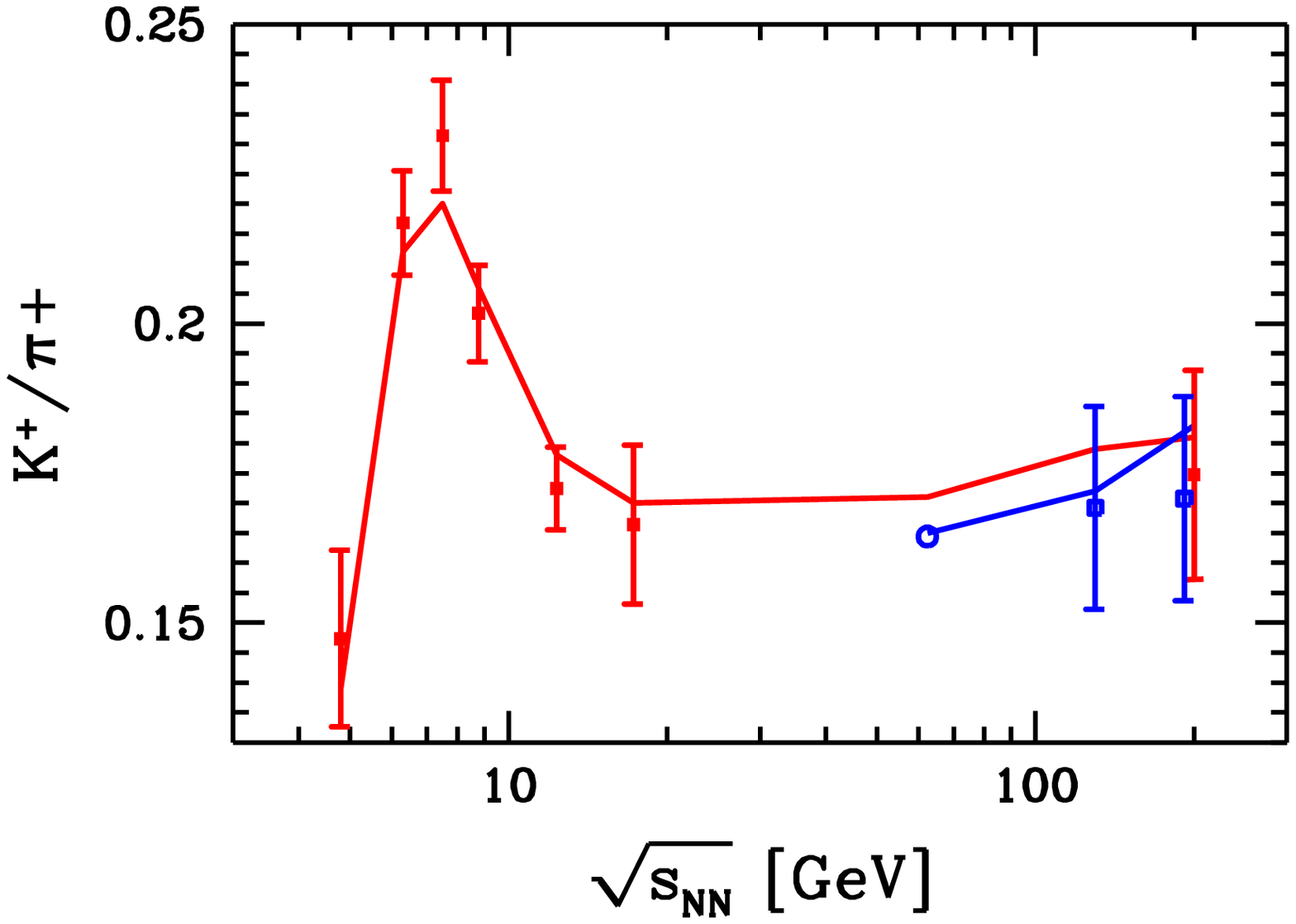}
\hspace*{-0.8cm}
\psfig{width=7.2cm, figure=\pathnow     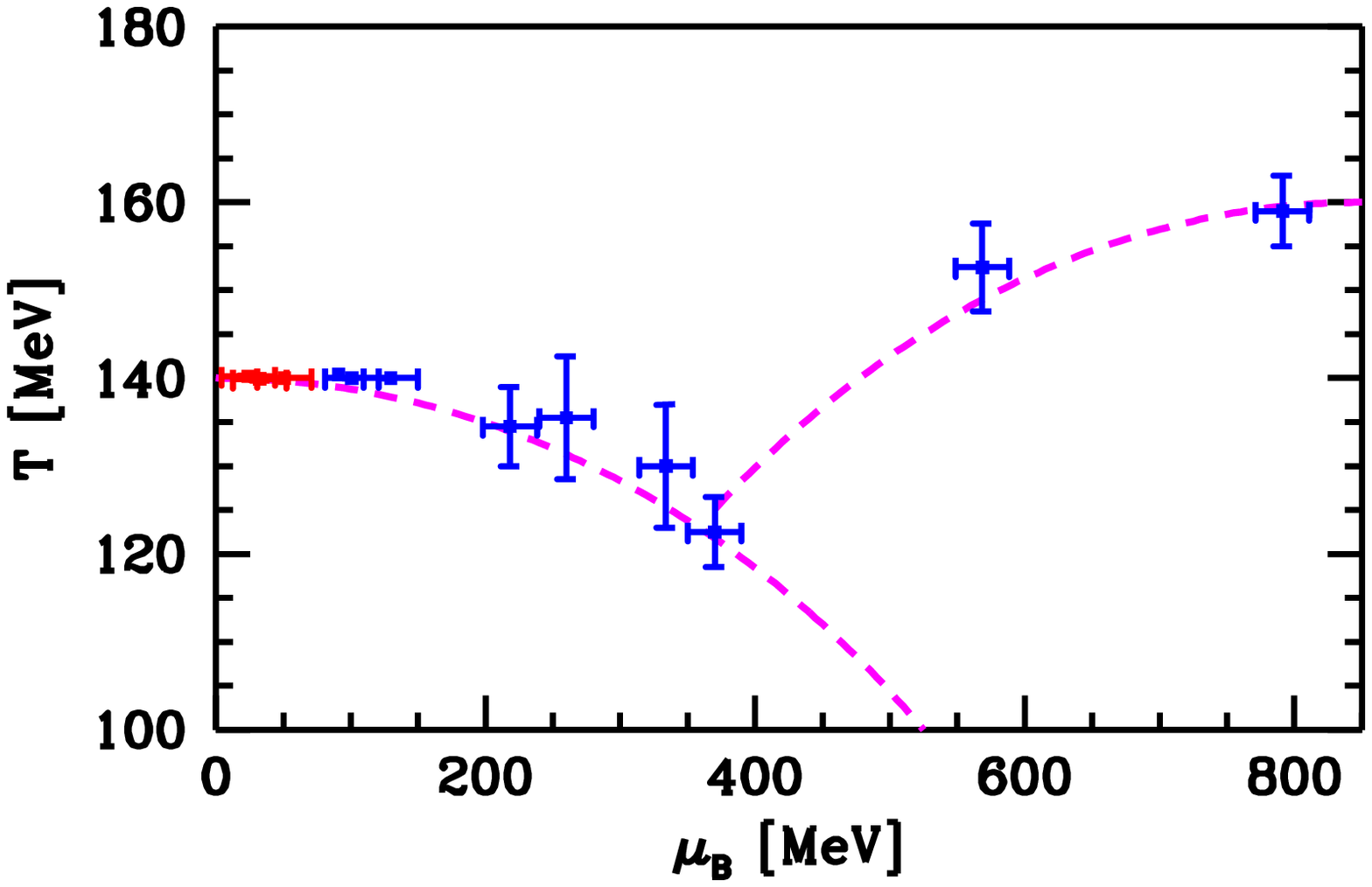}
}
\vskip -0.5cm
  \caption{\label{Kpisqrt}
On left  $K^+/\pi^+$ total yield ratio 
as function of $\sqrt{s_{\rm NN}}$ and its statistiacal 
hadronization model fit  employing 
chemical non-equilibrium.
At RHIC,  the central rapidity yield fit is also shown. On 
right, the fitted values of hadronization temperature and baryo-chemical
potential  are presented in the
$T$--$\mu_{\rm B}$ plane. The highest $\mu_{\rm B}$ entries on right
correspond to the lowest $\sqrt{s_{\rm NN}}$ 
values  on left~\cite{Letessier:2005qe}.
}\vskip -.6cm
\end{figure}

  This effect is  due to  
a rather sudden modification of   chemical conditions in the dense
matter fireball: the 
rapid rise in strangeness $\bar s$ production below,  and  a   rise in 
the anti-quark $\bar d$  yield above the `horn'. 
The measured ${\rm K}^+/{\pi}^+$ ratios are fitted well as is shown by the 
continuous line in  \rf{Kpisqrt}, when allowing for chemical non-equilibrium
at hadronization.  Both total yield ratios and the central rapidity ratios (for RHIC)
have been studied as shown on left in \rf{Kpisqrt}.
 
 A fit to this 
horn generates an `inverted horn'    in the  $T$--$\mu_{\rm B}$ plane,  
 see   right hand side of  \rf{Kpisqrt} . The RHIC $dN/dy$ results are to outer left.
They are followed by RHIC and SPS $N_{4\pi}$ results. The dip
corresponds to the 30 and 40 $A$GeV SPS results. The top right is 
the lowest  20 $A$GeV SPS and top 11.6 $A$GeV AGS energy range. 
To guide the eye, we have
added two lines connecting the fit results.  We see that the  
chemical freeze-out temperature 
$T$ rises for the two lowest reaction energies  11.6 and 20 $A$ GeV
 to near the Hagedorn temperature,  $T=160$ MeV, of boiling hadron  
matter. 
 
Such a non-trivial  hadronization boundary, 
in the $T$--$\mu_{\rm B}$ plane,   is the result
of a complex interplay between the dynamics of 
heavy ion reaction, and the properties of both phases of
matter, the inside of the fireball,  and the hadron phase we
observe.  The dynamical effect,  
capable to shift the location in temperature of 
the expected phase boundary is due to 
the expansion dynamics of the fireball, see section \ref{exp},
and effects of chemical non-equilibrium, see section \ref{Tphase}
for full discussion. However, the behavior seen at lower reaction energies 
maybe in part due to presence of effectively massive quarks, which do  not
chemically equilibrate.

\section{Highlights}
Our objective has been to show that there is a good reason to 
expect that the behavior of the QGP formed in heavy ion collisions
can deviate in a significant manner from expectations formed in 
study of equilibrium thermal QCD matter. We have described two main
effects, the chemical non-equilibrium of strange quarks, and the 
pressure of flowing color charge acting on the vacuum, which, 
in our opinion, are at current level of knowledge relevant to the
issues considered. 
Of most theoretical relevance and interest are the 
implications of non-equilibrium hadronization on the possible
change in the location and {\it nature} of the 
 phase boundary.

We have further argued that strangeness, and entropy,
are  well developed  tools
allowing  the detailed  study of hot  QGP phase. 
We have shown that it is possible to describe the `horn' in 
the $K^+/\pi^+$ hadron ratio within the chemical non-equilibrium 
statistical hadronization model.  The appearance
of this structure is related to a rapid change in the properties 
of the hadronizing matter.


\section*{Acknowledgments}
Work supported  by a 
grant from: the U.S. Department of
Energy  {\small DE-FG02-04ER4131}.\\ 
LPTHE, Univ.\,Paris 6 et 7 is: Unit\'e mixte de Recherche du {\small CNRS, UMR7589}. 


\vfill\eject

\begin{thebibliography}{19}
\bibitem{Peikert:1998jz}
  A.~Peikert, F.~Karsch, E.~Laermann and B.~Sturm,
  Nucl.\ Phys.\ Proc.\ Suppl.\  {\bf 73}, 468 (1999).


\bibitem{Fodor:2004nz}
Z.~Fodor and S.~D.~Katz,
JHEP {\bf 0404}, 050 (2004).


\bibitem{Karsch:2003jg}
  F.~Karsch and E.~Laermann,
In R.C. Hwa,  et al.: Quark gluon plasma III (2004) pp 1-59\ (World
Scientific, Singapore).
 

\bibitem{Bernard:2004je}
  C.~Bernard {\it et al.}  [MILC Collaboration],
  Phys.\ Rev.\ D {\bf 71}, 034504 (2005).
 
  

\bibitem{LHCPred}
  J.~Rafelski and J.~Letessier,
  arXiv:hep-ph/0506140, Euro. J. Phys. C in press. 


\bibitem{cool}
  J.~Letessier, J.~Rafelski and A.~Tounsi,
  Phys.\ Rev.\ C {\bf 50}, 406 (1994).


\bibitem{Letessier:2005qe}
  J.~Letessier and J.~Rafelski,
  ``Hadron production and phase changes in relativistic heavy ion collisions,''
  arXiv:nucl-th/0504028.
 

\bibitem{CUP}
  J.~Letessier and J.~Rafelski,
  {\it Hadrons and quark--gluon plasma}
Cambridge Monogr. Part. Phys. Nucl. Phys. Cosmol. {\bf 18} 1-397 (2002). 
Available  on line
at http://site.ebrary.com/pub/cambridge\-press/Doc?isbn=0521385369 


\bibitem{Rafelski:1999gq}
  J.~Rafelski and J.~Letessier,
  Phys.\ Lett.\ B {\bf 469}, 12 (1999).

\bibitem{Biro:1993qt}
  T.~S.~Biro {\it et al.},
  Phys.\ Rev.\ C {\bf 48}, 1275 (1993).

\bibitem{Pal:2001fz}
  D.~Pal, A.~Sen, M.~G.~Mustafa and D.~K.~Srivastava,
  Phys.\ Rev.\ C {\bf 65}, 034901 (2002).
 
\bibitem{He:2004df}
  Z.~J.~He {\it et al.}, 
  Phys.\ Rev.\ C {\bf 69}, 034906 (2004).


\bibitem{impact}
J.~Letessier, A.~Tounsi and J.~Rafelski,
Phys.\ Lett.\ B {\bf 389} (1996) 586.


\bibitem{Letessier:2003uj}
  J.~Letessier and J.~Rafelski,
  Phys.\ Rev.\ C {\bf 67}, 031902 (2003).



\bibitem{Rafelski:2000by}
  J.~Rafelski and J.~Letessier,
  Phys.\ Rev.\ Lett.\  {\bf 85}, 4695 (2000).

\bibitem{Csorgo:2002kt}
  T.~Csorgo and J.~Zimanyi,
  Heavy Ion Phys.\  {\bf 17}, 281 (2003).

\bibitem{Letessier:1992xd}
J.~Letessier {\it et al.}, 
Phys.\ Rev.\ Lett.\  {\bf 70}, 3530 (1993)
[arXiv:hep-ph/9711349].

 

\bibitem{RM82}
{J. Rafelski and B. Muller}, 
\newblock {\it Phys. Rev. Lett} {\bf 48}, 1066 (1982); {\bf 56}, 2334E (1986).
 

\bibitem{Rafelski:2004dp}
  J.~Rafelski, J.~Letessier and G.~Torrieri,
  Phys.\ Rev.\ C {\bf 72}, 024905 (2005).



\bibitem{Glendenning:1984ta}
  N.~K.~Glendenning and J.~Rafelski,
  Phys.\ Rev.\ C {\bf 31}, 823 (1985).


\bibitem{Gaz}
M.~Gazdzicki {\it et al.}  [NA49 Collaboration],
J.\ Phys.\ G {\bf 30}, S701 (2004).


\end{thebibliography}
\end{document}